\documentclass[pre,twocolumn,superscriptaddress,amsmath,amssymb,showpacs,showkeys]{revtex4}
\usepackage{dcolumn} 
\usepackage{epsfig}

\newcommand{\rmd}{\mathrm{d}}
\newcommand{\ket}[1]{|#1\rangle}
\newcommand{\bra}[1]{\langle#1|}
\newcommand{\braket}[1]{\langle#1\rangle}

\newcommand{\tr}{\operatorname{trace}}
\newcommand{\diag}{\operatorname{diag}}

\begin{document}

\title{Open source FCI code for quantum dots and effective interactions}
\date{October 15. 2008}

\author{Simen Kvaal}
\email{simen.kvaal@cma.uio.no}
\affiliation{Centre of Mathematics for Applications, University of
  Oslo, N-0316 Oslo, Norway}

\begin{abstract}
  We describe \textsc{OpenFCI}, an open source implementation of the
  full configuration-interaction method (FCI) for two-dimensional
  quantum dots with optional use of effective renormalized
  interactions. The code is written in C++ and is available under the
  Gnu General Public License. The code and core libraries are well
  documented and structured in a way such that customizations and
  generalizations to other systems and numerical methods are easy
  tasks. As examples we provide a matrix element tabulation program
  and an implementation of a simple model from nuclear physics, in
  addition to the quantum dot application itself.
\end{abstract}



\maketitle

\section{Introduction}
\label{sec:introduction}

Quantum dots, nanometre-scale semiconductor devices confining a
varying number of electrons, have been studied intensely in the last
two decades. Quantum dots are fabricated using essentially macroscopic
tools, for example etching techniques, but the resulting confinement
allows for quantum mechanical behaviour of the electrons. Many of the
parameters are directly controllable, thereby justifying the term
``artificial atoms'' or ``designer atoms''. These considerations
explain the immense research activity on these systems. For a general
introduction, see Ref.~\cite{Reimann2002} and references therein.
 
A very common model is that of a parabolic quantum dot, in which $N$
electrons are confined in an isotropic harmonic oscillator potential
in $d$ spatial dimensions, where $d$ is determined by the
semi-conductor environment. Electronic structure calculations on the
parabolic dot and similar systems are often carried out using the full
configuration-interaction method (FCI), also called exact
diagonalization \cite{Reimann2002}. The
Hamiltonian is then projected onto a finite-dimensional subspace of
the $N$-electron Hilbert space and diagonalized. Care is taken in
order to exploit dynamical and discrete symmetries of the exact
problem, such as conservation of angular momentum and total electron
spin, in order to block-diagonalize the Hamiltonian matrix and reduce
the computational complexity.

In this article, we describe \textsc{OpenFCI}, a recently developed
open source C++ code implementing the FCI method for quantum dots
\cite{OpenFCI}. The code has a generic framework in the shape of
library functions, thereby allowing easy customization and extension
to other systems and methods, e.g., three-dimensional quantum dots or
the nuclear no-core shell model.

\textsc{OpenFCI} implements a renormalization of the two-body
interactions, a technique widely used in nuclear no-core shell model
calculations. This allows for accelerated convergence with respect to
Slater determinant basis size \cite{Kvaal2008,Navratil2000}. To the
author's knowledge, no other available code provides such effective
interactions for quantum dot systems. The code can be easily modified
to create effective interactions for almost many-body problem using a
harmonic oscillator basis.

The code is developed in a Linux environment using Gnu C++, and is
readily portable to other environments and compilers. The Fortran 77
libraries \textsc{Lapack} and \textsc{Arpack} are required, but as
these are available on a wide range of platforms, portability should
not be affected. \textsc{OpenFCI} is released under the Gnu General
Public License (Gnu GPL) \cite{Gnu} and is documented using Doxygen
\cite{Doxygen}. As an open source project, the code can freely be used
and modified.

The article is organized as follows: In Section \ref{sec:fci-method},
the FCI method is introduced in the context of the parabolic quantum
dot, where we also discuss the reduction of the Hamiltonian matrix by
means of commuting operators and configurational state functions. In
Section \ref{sec:effective-int} we discuss the effective two-body
interaction. As the technique is likely to be unfamiliar to most
readers outside the nuclear physics community, this is done in some
detail. In Section \ref{sec:code} we discuss the organization and use
of \textsc{OpenFCI}. We also give some results from example runs, and
in particular an analytically solvable non-trivial model due to
Johnson and Payne is considered \cite{Johnson1991}, where the only
modification of the parabolic quantum dot is the interaction. Finally,
we conclude our article in Section \ref{sec:conclusion}.

Two appendices have been provided, Appendix \ref{sec:com} detailing
the heavily-used centre-of-mass transformation and Appendix
\ref{sec:radial-method} discussing the exact numerical solution of the
two-electron quantum dot needed for the effective interaction scheme.

\section{FCI method}
\label{sec:fci-method}

\subsection{Hamiltonian in occupation number formalism}
\label{sec:occupation-number-formalism}

We consider $N$ electrons trapped in an isotropic harmonic oscillator
potential in $d$ spatial dimensions. The electrons interact via the
Coulomb potential given by $U(r_{ij}) = \lambda / r_{ij}$, where
$r_{ij} = \| \vec{r}_i - \vec{r}_j \|$ is the inter-particle distance
and $\lambda$ is a constant. The
quantum dot Hamiltonian then reads
\begin{equation} 
  H := \sum_{i=1}^N H_0(i) + \sum_{i<j}^N U(r_{ij}), 
  \label{eq:H}
\end{equation}
where the second sum runs over all pairs $1\leq i<j\leq N$, and 
where $H_0(i)$ is the
one-body Hamiltonian defined by
\[ H_0(i) := -\frac{1}{2}\nabla_i^2 +
\frac{1}{2}\|\vec{r}_i\|^2. \]
The interaction strength $\lambda$ is
given by
\begin{equation}
  \lambda = \sqrt{\frac{m^*}{\omega\hbar^3}} \frac{1}{\epsilon} \frac{e^2}{4\pi\epsilon_0},
  \label{eq:lambda}
\end{equation}
where $\epsilon$ is the dielectric constant of the semiconductor bulk,
$e^2/4\pi\epsilon_0 \approx 1.440$ eV$\cdot$nm, and $\omega = \hbar/m^* a^2$, $a$ being
the trap size and length unit, and $m^*$ being the effective electron
mass. Typical values for GaAs quantum dots are $\epsilon=12.3$, $m^* =
0.067$ electron masses, and $a = 20$ nm, yielding $\lambda =
2.059$. The energy unit is $\hbar\omega$, in this case $\hbar\omega =
2.84$ meV.

Choosing a complete set $\{\phi_\alpha(x)\}_{\alpha\in A}$ of
single-particle orbitals (where $x=(\vec{r},s)$ denotes both spatial
and spin degrees of freedom, and $\alpha=(a,\sigma)$ denotes both
generic spatial quantum numbers $a$ and spin projection quantum numbers
$\sigma=\pm 1$), $H$ can be written in occupation number
form as
\begin{equation} 
  H = \sum_{a,b} \sum_\sigma h^a_b a^\dag_{a,\sigma}a_{b,\sigma} 
  + \frac{1}{2}\sum_{abcd}\sum_{\sigma\tau} u^{ab}_{cd}
  a^\dag_{a,\sigma}a^\dag_{b,\tau}a_{d,\tau}a_{c,\sigma},
\label{eq:H-occ}
\end{equation}
where $a^\dag_\alpha$ ($a_\alpha$) creates (destroys) a particle in
the orbital $\phi_\alpha(x)$. These operators obey the usual
anti-commutation relations
\begin{equation}
  \{ a_\alpha, a^\dag_\beta \} = \delta_{\alpha,\beta}, \quad \{
  a_\alpha, a_\beta \} = 0. 
  \label{eq:anticommutators}
\end{equation}
For a review of second quantization and
occupation number formalism, see for example
Ref.~\cite{Raimes1972}. The single-particle orbitals are chosen on the form
\[ \phi_{(a,\sigma)}(x) := \varphi_a(\vec{r})\chi_\sigma(s), \] where
$\{\varphi_a(\vec{r})\}$ are spinless orbitals and $\chi_\sigma(s) =
\delta_{\sigma,s}$ are spinor basis functions corresponding to the
eigenstates of the spin-projection operator $S_z$ with eigenvalues
$\sigma/2$. 

It is important, that since the single-particle orbitals
$\{\varphi_a(\vec{r})\}_{a \in A}$ are denumerable, we may choose an
ordering on the set $A$, such that $A$ can in fact be identified with a
range of integers, $A \approxeq \{0,1,2,\cdots, L/2\}$. In most \emph{ab
  initio} systems $L$ is infinite, since the Hilbert space is infinite-dimensional. Similarly, $\alpha=(a,+1)$ is identified with even
integers, and $\alpha=(a,-1)$ with odd integers, creating an ordering
of the single-particle orbitals $\phi_\alpha(x)$, and $\alpha$ is
identified with an integer $0 \leq I(\alpha) \leq L$.

The single-particle matrix elements $h^a_b$
and the two-particle elements $u^{ab}_{cd}$ are defined by
\[ h^a_b := \langle \varphi_a | H_0 | \varphi_b \rangle = \int
\overline{\varphi_a(\vec{r})}H_0\varphi_b(\vec{r}) \; \rmd^d r, \]
and
\begin{eqnarray}
  u^{ab}_{cd} &:=& \langle \varphi_a \varphi_b | U(\vec{r}_{12}) |
  \varphi_c \varphi_d \rangle \notag \\ &=& \lambda \int \overline{\varphi_a(\vec{r}_1)}
  \overline{\varphi_b(\vec{r}_2)} \frac{1}{r_{12}} \varphi_c(\vec{r}_1)
  \varphi_d(\vec{r}_2) \; \rmd^d r_1 \rmd^d r_2, \label{eq:inter-matrix}
\end{eqnarray}
respectively. 

The spatial orbitals $\varphi_a(\vec{r})$ are usually chosen as
eigenfunctions of $H_0$, so that $h^a_b = \delta_{a,b}\epsilon_a$.

The basis functions for $N$-particle Hilbert space are Slater determinants
$|\Phi_{\alpha_1,\alpha_2,\cdots,\alpha_N} \rangle$ defined by
\[ |\Phi_{\alpha_1,\cdots\alpha_N}\rangle :=
a^\dag_{\alpha_1}a^\dag_{\alpha_2}\cdots a^\dag_{\alpha_N}
|-\rangle, \]
where $|-\rangle$ is the zero-particle vacuum. In terms of
single-particle orbitals, the spatial representation is
\[ \Phi_{\alpha_1,\cdots,\alpha_N}(x_1,\cdots,x_N) =
\frac{1}{\sqrt{N!}}\sum_{p\in S_N} (-)^{|p|} \prod_{i=1}^N
\phi_{\alpha_{p(i)}}(x_i), \] where $S_N$ is the group of
permutations of $N$ symbols. The Slater determinants are
anti-symmetric with respect to permutations of both $x_i$ and
$\alpha_i$, so that the orbital numbers $\alpha_i$ must all be
distinct to give a nonzero function. Each orbital is then occupied by
at most one particle. Moreover, for a given set $\{\alpha_i\}_{i=1}^N$
of orbitals, one can create $N!$ distinct Slater determinants that are
linearly dependent. In order to remove this ambiguity, we choose only
orbital numbers such that $I(\alpha_i)<I(\alpha_j)$ whenever $i<j$.

It follows, that there is a natural one-to-one correspondence between
Slater determinants with $N$ particles and integers $b$ whose binary
representations have $N$ bits set. (If $|A|=L < \infty$, the integers
are limited to $0\leq b < 2^L$.) Each bit position $k$ corresponds to
an orbital $\phi_\alpha(x)$ through $k=I(\alpha)$, and the bit is set
if the orbital is occupied. Creating and destroying particles in
$|\Phi_{\alpha_1,\cdots,\alpha_N}\rangle$ simply amounts to setting or
clearing bits (possibly obtaining the zero-vector if a particle is
destroyed or created twice in the same orbital), keeping track of the
possible sign change arising from bringing the set $\{\alpha_i\}$ on
ordered form using Eqn.~(\ref{eq:anticommutators}). Note that the
vacuum $\ket{-}$ corresponds to $b=0$, which is \emph{not} the zero
vector, but the single state with zero particles.

\subsection{Model spaces}
\label{sec:model-spaces}

The FCI calculations are  done in a finite-dimensional subspace
$\mathcal{P}$ of the $N$-particle Hilbert space, called the model space. The model
space has a basis $\mathcal{B}$ of Slater determinants, and
$\mathcal{P}$ has the orthogonal projector $P$ given by
\begin{equation}
  P := \sum_{\ket{\Phi_b}\in\mathcal{B}} \ket{\Phi_b}\bra{\Phi_b}.
  \label{eq:projector}
\end{equation}
The configuration-interaction method in general now amounts to
diagonalizing (in the sense of finding a few of the lowest eigenvalues
of) the, in general, large and sparse matrix $PHP$. The only
approximation we have made is the truncation of the $N$-particle
Hilbert space.

\begin{figure}
\includegraphics{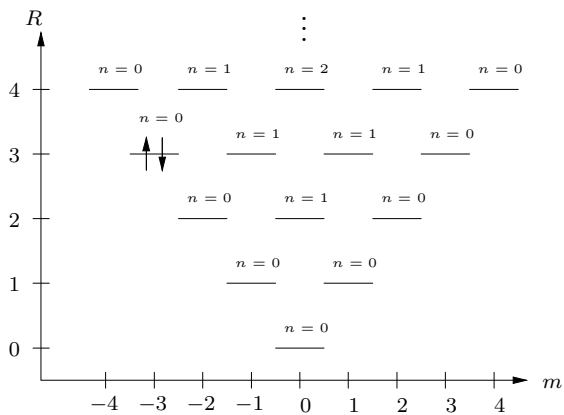}
\caption{Structure of single-particle orbitals of the two-dimensional
  harmonic oscillator. Angular momentum and shell number/energy on
  axes, and nodal quantum number $n$ at each orbital. Orbital $n=0$,
  $m=-3$ in shell $R=3$ is occupied by two electrons for
  illustration.\label{fig:shells}}
\end{figure}

The model space $\mathcal{P}$ is seen to be a function of the single
particle orbitals $\varphi_a(\vec{r})$, whom we choose to be the eigenfunctions
of $H_0$, i.e., harmonic oscillator eigenfunctions. These may be given
on several equivalent forms, but it is convenient to utilize rotational
symmetry of $H_0$ to create eigenfunctions of the projection
of the angular momentum $L_z$. In $d=2$ dimensions we obtain the
Fock-Darwin orbitals defined in polar coordinates by
\begin{equation} 
  \varphi_{n,m}(r,\theta) = \frac{1}{\sqrt{\pi}} e^{im\theta} r^{|m|}
  \tilde{L}_n^{|m|}(r^2)
 e^{-r^2/2}. \label{eq:fock-darwin-def} 
\end{equation}
Here $\tilde{L}_n^k(x)=(-1)^n[n!/(n+|m|)!]^{1/2}L_n^k(x)$ is the
normalized generalized Laguerre polynomial. The factor $(-1)^n$ is for
convenience, see Appendix \ref{sec:com-cartesian}. The harmonic
oscillator energy is $2n+|m|+1$ and the eigenvalue of
$L_z=-i\partial/\partial_\theta$ is $m$. All eigenfunctions with the
same energy
$2n+|m|+1=:R+1$ span a single-particle \emph{shell}.
The single-particle orbitals are illustrated in Fig.~\ref{fig:shells}. 

For a Slater determinant $|\Phi_{\alpha_1,\cdots,\alpha_N}\rangle$, we
have
\[ \sum_{i=1}^N H_0(i) |\Phi_{\alpha_1,\cdots,\alpha_N}\rangle =
E^0_{\alpha_1,\cdots,\alpha_N}|\Phi_{\alpha_1,\cdots,\alpha_N}\rangle \]
with 
\[ E^0_{\alpha_1,\cdots,\alpha_N} := \sum_{i=1}^N (R_i + 1), \]
where $R_i = 2n_i + |m_i|$, and
\[ \sum_{i=1}^N L_z(i) |\Phi_{\alpha_1,\cdots,\alpha_N}\rangle =
M |\Phi_{\alpha_1,\cdots,\alpha_N}\rangle, \]
where $M = \sum_{i=1}^N m_i$.

To complete our definition of $\mathcal{P}$, we let
\begin{equation}
  \mathcal{B} = \mathcal{B}_R = \left\{ \ket{\Phi_{\alpha_1,\cdots,\alpha_N}} \;:\;
    \sum_{i=1}^N R_i \leq R \right\}, 
  \label{eq:energy-cut}
\end{equation}
where $R$ is called the energy cut, for obvious reasons. As
$R\rightarrow\infty$, the whole Hilbert space is spanned, and the
eigenpairs of $PHP$ converge to those of $H$.

\subsection{Configurational state functions and block diagonality}
\label{sec:csf}

In order to reduce the complexity of the computations, we need to
exploit symmetries of $H$. First of all, $[H,L_z]=0$, and it is
obvious that also $[H,S_z]=0$, where the spin projection operator
$S_z$ is given by
\[ 
S_z := \frac{1}{2}\sum_{a,\sigma}\sigma a^\dag_{a,\sigma}
a_{a,\sigma}. 
\] 
The Slater determinants are eigenvectors of both $L_z$ and $S_z$ with
eigenvalues $M$ and $s_z=\sum_{i=1}^N\sigma_i/2$, respectively. We
obtain a natural splitting of the model space $\mathcal{P}$ into
subspaces with constant angular momentum $M$ and spin projection
$s_z$, viz,
\[ \mathcal{P} = \bigoplus_{M,s_z} \mathcal{P}_{M,s_z}, \quad P = \sum_M\sum_{s_z} P_{M,s_z}. \] 
The diagonalization of $H$ can thus be done within each space
$\mathcal{P}_{M,s_z}$ separately, amounting to diagonalizing
individual blocks $P_{M,s_z} H P_{M,s_z}$.

The Hamiltonian (\ref{eq:H-occ}) also commutes with total
electron spin $S^2$, $[S^2,S_z]=0$, given by
\[ S^2 := S_z^2 + \frac{1}{2}(S_+S_- + S_-S_+), \]
with
\[ S_\pm := \sum_a a^\dag_{a_\pm} a_{a_\mp}, \] so that a common basis
for $S_z$ and $S^2$ would lead to even smaller matrix blocks.

The eigenvalues of $S^2$ are on the form $s(s+1)$, where $0 \leq
2s\leq N$ is an odd (even) integer for odd (even) $N$. For a joint
eigenfunction of $S_z$ and $S^2$, called a configurational state
function (CSF), $|s_z|\leq s$. The Slater determinants are, however,
not eigenfunctions of $S^2$, but such can be constructed by taking
linear combinations of a small number Slater determinants. For details
on this algorithm, see Ref.~\cite{Rontani2006}. Suffice it to say
here, that $S^2$ only couples Slater determinants with identical sets
of doubly occupied orbitals (meaning that $\phi_{(a,+)}$ and
$\phi_{(a,-)}$ are both occupied, as in Figure \ref{fig:shells}) and
singly occupied orbitals (meaning that only one of $\phi_{(a,+)}$ and
$\phi_{(a,-)}$ are occupied). It is easy to see that $S^2$ does not
couple Slater determinants in $\mathcal{P}_{M,s_z}$ to another
$\mathcal{P}_{M',s_z'}$. Thus, we obtain the splitting
\[ \mathcal{P}_{M,s_z} = \bigoplus_{s} \mathcal{P}_{M,s_z,s}. \]
We stress that all the mentioned operators commute with each
other, viz,
\[ [H,\Omega_i] = [\Omega_i,\Omega_j] = 0, \]
with $\Omega_i \in \{ L_z, S_z, S^2 \}$. If a modified problem breaks,
say, rotational symmetry, such that $[H,L_z]\neq 0$, we may still
split the model space into to the eigenspaces of $S_z$ and $S^2$.

\subsection{Matrix elements of Coulomb interaction}
\label{sec:coulomb-matrix}

The remaining ingredient in the FCI method is the Coulomb matrix
elements $u^{ab}_{cd}$ defined in Eqn.~(\ref{eq:inter-matrix}). These
can be calculated by first expanding $L_n^k(x)$ in powers of $x$ using
\[ L_n^k(x) \equiv \sum_{m=0}^n (-1)^m
\frac{(n+k)!}{(n-m)!(k+m)!m!}x^m, \] and evaluating the resulting
integral term-by-term by analytical methods \cite{Anisimovas1998}.
The resulting expression is a seven-fold nested sum, which can be
quite time-consuming, especially if a large number of Fock-Darwin
orbitals occurs in the basis $\mathcal{B}$. Moreover, the terms
are fractions of factorials with alternating signs, which is a
potential source of loss of numerical precision. 

We therefore opt for a more indirect approach, giving a procedure
applicable to a wide range of potentials $U(r_{12})$ in addition to
the Coulomb potential. Moreover, it can be generalized to arbitrary
spatial dimensions $d$. The approach is based on directly transforming
the product functions $\varphi_a(\vec{r}_1)\varphi_b(\vec{r}_2)$ to
the centre-of-mass system, where the interaction $U(r_{12})$ only acts
on the relative coordinate, and then transforming back to the lab
system. This reduces the computational cost to a doubly nested sum, as
well as the pre-computation of the centre-of-mass transformation and
the relative coordinate interaction matrix. Both can be done exactly
using Gaussian quadrature. The transformations to and from the centre
of mass frame are unitary transformations, which are stable and will
not magnify round-off errors.

In Appendix \ref{sec:com} we provide the details of the
centre-of-mass transformation. One then obtains the following
prescription for the interaction matrix elements $u^{ab}_{cd}$: Let $a
= (\mu_1,\nu_1)$, $b = (\mu_2, \nu_2)$, $c = (\mu_3,\nu_3)$, and $d =
(\mu_4,\nu_4)$ be the circular quantum number equivalents of the usual
polar coordinate quantum numbers $n_i$ and $m_i$. Due to conservation
of angular momentum, we assume $m_1 + m_2 = m_3 + m_4$; otherwise, the
matrix element $u^{ab}_{cd} = 0$. Define $M=\mu_1+\mu_2$,
$M'=\mu_3+\mu_4$, $N=\nu_1+\nu_2$, and $N'=\nu_3+\nu_4$.
Since $u^{ab}_{cd}$ is linear in $\lambda$, we set
$\lambda=1$ without loss of generality. Now, 
\begin{eqnarray}
  u^{ab}_{cd} &=& \sum_{p = p_0}^{M} 
  T^{(M)}_{p,\mu_2} T^{(M')}_{p',\mu_4} \sum_{q = q_0}^{N} 
  T^{(N)}_{q,\nu_2} T^{(N')}_{q',\nu_4} C^{|p-q|}_{n,n+s}, \label{eq:interaction-elems}
\end{eqnarray}
where $n = \min(p,q)$, $s = M'-M$, $p' = p + M'-M$, $q'=q +
N'-N$. Moreover, $p_0 = \max(M'-M, 0)$ and $q_0 = \max(N'-N, 0)$.

Here, $T^{(N)}$ are centre-of-mass transformation coefficients defined
in Appendix \ref{sec:com}, while the relative coordinate interaction
matrix elements $C^{|m|}_{n,n'}$, $n,n'\geq 0$, are defined by
\begin{eqnarray} 
  C^{|m|}_{n,n'} &:=&
  \braket{\varphi_{n,m}(r,\theta)|U(\sqrt{2}r)|\varphi_{n',m}(r,\theta)}
  \notag \\
&=&
2\int_{0}^\infty
r^{2|m|} \tilde{L}_n^{|m|}(r^2) 
\tilde{L}_{n'}^{|m|}(r^2) U(\sqrt{2}r) e^{-r^2} r \rmd r. \label{eq:coulomb-integral}
\end{eqnarray}
Depending on $U(r_{12})$, the integral is best computed using
generalized half-range Hermite quadrature (see Appendix
\ref{sec:radial-method} and Ref.\cite{Ball2002}) or Gauss-Hermite
quadrature. Weights and abscissa for quadratures are conveniently
computed using the Golub-Welsch algorithm \cite{Golub1969}, which only
depends on the ability to compute the coefficients of the three-term
recursion relation for the polynomial class in question, as well as
diagonalizing a symmetric tri-diagonal matrix.

Let $p(r)$ be a
polynomial, and let $\alpha<2$ and $\beta$ be non-negative
constants. Then
\begin{equation}
  U(r_{12}) = r_{12}^\alpha p(r_{12}) e^{-\beta r_{12}^2} 
  \label{eq:pot1}
\end{equation}
admit exact evaluations using generalized half-range Gauss-Hermite
quadrature. The Coulomb potential, Gaussian potentials, and the
parabolic interaction $-\lambda r_{12}^2/2$ of the analytically
solvable model treated in Sec.~\ref{sec:results} belong to this class
of potentials.

In the case of $\alpha = 1$ and $p(r) = q(r^2)$ (i.e., an even
polynomial), the integral is more convenient to evaluate using
standard Gauss-Hermite quadrature. The Coulomb interaction falls into
this class.

Of course, one may let $p(r)$ be a non-polynomial function as well and
still obtain very good results, as long as $p(r)$ is well approximated
with a polynomial, e.g., is smooth.

\section{Effective interactions}
\label{sec:effective-int}

\subsection{Motivation}

The FCI calculations converge relatively slowly as function of the
model space parameter $R$ \cite{Kvaal2008}, as the error $\Delta E$ in
the eigenvalue behaves like $o(R^{-k})$ in general, where
$k=O(1)$. This behaviour comes from the singular nature of the
Coulomb interaction. 

In Ref.~\cite{Kvaal2008}, numerical results using an effective
interaction were presented. This method is widely used in no-core
shell model calculations in nuclear physics, where the nucleon-nucleon
interaction is basically unknown but highly singular
\cite{Navratil2000}. This so-called sub-cluster effective interaction
scheme
replaces the Coulomb interaction (or another interaction) $U(r_{ij}) =
\lambda/r_{ij}$ with a renormalized interaction $\tilde{U}(i,j)$
obtained by a unitary transformation of the two-body Hamiltonian that
decouples the model space $\mathcal{P}$ and its complement
\cite{Kvaal2008a}. Therefore, the two-body problem becomes
\emph{exact} in a \emph{finite} number of harmonic oscillator
shells. Loosely speaking, the effective interaction incorporates
information about the interaction's action \emph{outside} the model
space. In general, $\tilde{U}(i,j)$ is non-linear in $\lambda$ and not
a local potential.

Using the renormalized $\tilde{U}(i,j)$, the many-body system does not
become exact, of course, but $\tilde{U}(i,j)$ will
perform better than the bare interaction in this setting as well. To
the author's knowledge, there exists no rigorous mathematical treatment
with respect to this, but it has nevertheless enjoyed great success in
the nuclear physics community
\cite{Barrett2006,Navratil2000,HjorthJensen1995}, and our numerical
experiments unambiguously demonstrate that the convergence of the FCI
method is indeed improved drastically 
\cite{Kvaal2008}, especially for $N \leq 4$ particles. We stress that
the cost of producing $\tilde{U}(i,j)$ is very small compared to the
remaining calculations.

\subsection{Unitary transformation of two-body Hamiltonian}
\label{sec:unitary-transformation}

We now describe the unitary transformation of the two-body Hamiltonian
(i.e., Eqn.~(\ref{eq:H}) or (\ref{eq:H-occ}) with $N=2$) that de-couples $\mathcal{P}$ and
its complement. This approach dates back as far as 1929, when Van Vleck introduced
such a generic unitary transformation to de-couple the model space to
first order in the interaction \cite{VanVleck1929,Kemble1937}.

Let $P$ be given by Eqn.~(\ref{eq:projector}), and let
$D=\dim(\mathcal{P})$. 
The idea is to find a unitary transformation
$\mathcal{H}=Z^\dag H Z$ of $H$ 
such that 
\[ 
(1-P)\mathcal{H}P = 0, 
\] 
i.e., $\mathcal{H}$ is block
diagonal. This implies that $H_\text{eff}$ defined by
\[ 
H_\text{eff} := P\mathcal{H}P
\]
has eigenvalues identical to $D$ of those of the \emph{full operator}
$H$. Since $D$ is finite, $H_\text{eff}$ is called an effective
Hamiltonian.

Selecting $Z$ is equivalent to selecting a set of effective
eigenpairs $\{(E_k,\ket{\Psi_k^\text{eff}})\}_{k=1}^D$, where $E_k$ is
an eigenvalue of $H$ and $\{\ket{\Psi_k^\text{eff}}\}_{k=1}^D \subset
\mathcal{P}$ are the effective eigenvectors; an orthonormal basis for
$\mathcal{P}$. It is clear that $Z$ is not unique, since there are
many ways to pick $D$ eigenvalues of $H$, and for each such selection
any unitary $D\times D$ matrix would yield an eigenvector set.

However, some choices are more natural than others, since the
eigenvectors and eigenvalues are usually continuous functions of
$\lambda$. We then select the $D$ eigenvalues $E_k(\lambda)$ that develop
adiabatically from $\lambda = 0$. For the corresponding effective eigenvectors
$\ket{\Psi_k^\text{eff}(\lambda)}$, we choose the orthonormal set that
minimizes the distance to the exact eigenvectors $\{\ket{\Psi_k}\}_{k=1}^D$, i.e.,
\begin{equation} \{ \ket{\Psi_k^\text{eff}} \}_{k=1}^D := \underset{
    \{\ket{\Psi_k'}\}_{k=1}^D }{\operatorname{argmin}} \sum_{k=1}^D \|
  \ket{\Psi_k} - \ket{\Psi_k'}
  \|^2, \label{eq:minimize} \end{equation} where the minimization is
taken over orthonormal sets only. The effective eigenvectors also turn
out to be continuous functions of $\lambda$, so $H_\text{eff}$ will also be
continuous.

Let $U$ is the $D\times D$ matrix whose columns contain
$P\ket{\Psi_k}$ in the chosen basis, and let $V$ be the corresponding
matrix containing $\ket{\Psi_k^\text{eff}}$. Clearly, $V$ is unitary,
while $U$ only approximately so. Equation (\ref{eq:minimize}) can then
be written
\begin{equation}
  V := \underset{U'}{\operatorname{argmin}}\; \tr
[(U-U')(U-U')^\dag], \label{eq:minimize2} \end{equation}
where the minimum is taken over all unitary matrices. If $U$ has singular
value decomposition given by
\begin{equation}
  U = X\Sigma Y^\dag,
\end{equation}
the solution $V$ is given by
\begin{equation}
  V := X Y^\dag.
\end{equation}
If $E = \diag(E_1,\cdots,E_D)$ is the diagonal matrix whose elements
are the chosen eigenvalues, we have
\[ H_\text{eff} = V E V^\dag. \]
See Ref.~\cite{Kvaal2008a} for a
thorough discussion of the above prescription for $H_\text{eff}$.

Having computed the two-body $H_\text{eff}$,
we define the effective interaction $\tilde{U}(1,2)$ by
\[ \tilde{U}(1,2) := H_\text{eff}  - P \sum_{i=1}^2 H_0(i) P, \]
which gives meaning \emph{solely} in the model space.
In second quantization, 
\[ \tilde{U}(1,2) := \frac{1}{2} \sum_{abcd}\sum_{\sigma\tau} \tilde{u}^{ab}_{cd} a^\dag_{a\sigma}
a^\dag_{b\tau} a_{d\tau} a_{c\sigma}, \]
and the $N$-body $H_\text{eff}$ becomes (cf.~Eqn.~(\ref{eq:H}))
\[ H_\text{eff} = \sum_{i=1}^N H_0(i) + \sum_{i<j}^N
\tilde{U}(i,j), \]
with occupation number formalism form (cf.~Eqn.~(\ref{eq:H-occ}))
\begin{equation} 
  H_\text{eff} = \sum_{a,b} \sum_\sigma h^a_b a^\dag_{a,\sigma}a_{b,\sigma} 
  + \frac{1}{2}\sum_{abcd}\sum_{\sigma\tau} \tilde{u}^{ab}_{cd}
  a^\dag_{a,\sigma}a^\dag_{b,\tau}a_{d,\tau}a_{c,\sigma}.
\label{eq:Heff-occ}
\end{equation}
Now, $H_\text{eff}$ is well-defined in the space of $N$-body Slater
determinants where no pairs of occupied orbitals constitute a two-body state
outside the two-particle model space, since then the matrix element
$\tilde{u}^{ab}_{cd}$ would be undefined. A little thought shows us
that if $\tilde{U}(1,2)$ was computed in a two-body energy cut space
with parameter $R$, $H_\text{eff}$ is well-defined on the
many-body model space with the same cut $R$.

\subsection{A comment concerning the choice of model space}
\label{sec:model-space-comment}

The two-body problem is classically integrable, i.e., there exists
$2d-1$ constants of motion $\Omega_i$, such that their quantum mechanical
observables commute with $H$ and each other, viz,
\[ [H,\Omega_i] = [\Omega_i,\Omega_j] = 0, \quad \text{for all} \; i,j. \]
Indeed, the centre-of-mass harmonic oscillator $H_\text{C}$ defined in
Eqn.~(\ref{eq:com-separation}) below and the corresponding centre-of-mass
angular momentum provides two constants, while total angular
momentum $L_z$ provides a third.

Using the model space $\mathcal{P}$ defined by an energy cut, we have
\[ [P,\Omega_i] = 0 \]
as well, which is equivalent \cite{Kvaal2008a} to
\begin{equation}
  [H_\text{eff}, \Omega_i] = 0,
  \label{eq:Heff-commutator}
\end{equation}
so that $H_\text{eff}$ is integrable as well. In particular,
$\tilde{U}(1,2)$ is block-diagonal with respect to $\Omega_i$.

If we consider the commonly encountered model space $\mathcal{P}'$
defined by the Slater determinant basis $\mathcal{B}'$ given by
\[ \mathcal{B}' := \{ \ket{\Phi_{\alpha_1,\cdots,\alpha_N}} \;:\;
\max(R_i) \leq R \}
\]
instead of Eqn.~(\ref{eq:energy-cut}), we will have
\[ [P',H_\text{C}] \neq 0, \]
as is easily verified. Indeed, $\mathcal{P}'$ is not an invariant
subspace of the centre-of-mass transformation $T$ defined in Appendix
\ref{sec:com}. Thus, $[H'_\text{eff}, H_\text{C}] \neq 0$, so that the
centre-of-mass energy no longer is a constant of
motion! The symmetry-breaking of the effective Hamiltonian in this
case is problematic, since in the limit $\lambda\rightarrow 0$, the
exact eigenfunctions that develop adiabatically are not all either in
the model space or in the complement. The
adiabatic continuation of the eigenpairs starting out in $\mathcal{P}$
is thus not well-defined.

We comment, that the model space $\mathcal{P}'$ is often used in both no-core shell
model calculations and quantum dot calculations, but the effective
interaction becomes, in fact, ill-behaved in this case.

\subsection{Solution of the two-body problem}
\label{sec:exact-twobody}

What remains for the effective interaction, is the computation of the exact eigenpairs
$\{(E_k,\ket{\Psi_k})\}_{k=1}^D$. We must also solve the problem of
following eigenpairs adiabatically from $\lambda = 0$.

For the two-body Coulomb problem, analytical solutions are
available only for very special values for $\lambda$ \cite{Taut1993}. These
are useless for our purpose, so we must use numerical methods.

A direct application of the FCI method using Fock-Darwin orbitals with
a large $R'>R$ will converge slowly, and there
is no device in the method for following eigenvalues adiabatically. As
the eigenvalues may cross, selecting, e.g., the lowest eigenvalues
will not work in general.

For the two-body problem, the Pauli principle leads to a symmetric
spatial wave function for the singlet $s=0$ spin state, and an
anti-symmetric wave function for the triplet $s=1$ spin states. For
the spatial part, we exploit the integrability of the system as
follows. Define
centre-of-mass coordinates by
\begin{eqnarray*} 
  \vec{R} &:=& \frac{1}{\sqrt{2}}(\vec{r}_1 + \vec{r}_2) 
\end{eqnarray*}
and
\begin{eqnarray*} 
  \vec{r} &:=& \frac{1}{\sqrt{2}}(\vec{r}_1 - \vec{r}_2).
\end{eqnarray*}
Using these coordinates,
the two-body Hamiltonian becomes
\begin{eqnarray}
  H &=& H_0(\vec{R}) + \left[ H_0(\vec{r}) +
    U(\sqrt{2}r;\lambda)\right] \notag
  \\
  &=:& H_\text{C} + H_\text{rel} \label{eq:com-separation}
\end{eqnarray}
where $r_{12} = \sqrt{2}r := \sqrt{2}\|\vec{r}\|$. We have introduced
the parameter $\lambda$ explicitly in the potential in this equation. $H$ is
clearly separable, and the centre-of-mass coordinate Hamiltonian
$H_\text{C}$ is a trivial harmonic oscillator, while the relative
coordinate Hamiltonian can be written as
\[ H_\text{rel} := -\frac{1}{2}\nabla^2 + \frac{1}{2}r^2 +
U(\sqrt{2}r;\lambda), \]
where in polar coordinates $\vec{r} = (r\cos\theta,r\sin\theta)$ we have
\[ \nabla^2 =
\frac{1}{r}\frac{\partial}{\partial r}r\frac{\partial}{\partial r}
+ \frac{\partial^2}{\partial\theta^2}. \]
Applying separation of variables again, the eigenfunctions of $H_\text{rel}$
can be written
\[ \psi_{n,m}(\vec{r}) := \frac{e^{im\theta}}{\sqrt{2\pi}} u_{n,m}(r) \]
where $n$ is the nodal quantum number. $u_{n,m}(r)$ satisfies 
\begin{equation}
  K_{|m|} u_{n,m}(r) = \mu_{n,m} u_{n,m}(r)
    \label{eq:radial-eqn}
\end{equation}
where
\begin{equation}
  K_{|m|} :=  -\frac{1}{2r}\frac{\partial}{\partial
   r}r\frac{\partial}{\partial r}
    + \frac{m^2}{2r^2} + \frac{1}{2}r^2 +
    U(\sqrt{2}r; \lambda).
    \label{eq:radial-ham}
\end{equation} 

Equation (\ref{eq:radial-eqn}) is an eigenvalue problem in the Hilbert
space $L^2([0,\infty), r\rmd r)$, where the measure $r\rmd r$ is
induced by the polar coordinate transformation. Although it is natural
to try and solve the radial problem using Fock-Darwin orbitals, this
will converge slowly.  The
solution to this problem is to use a radial basis of generalized
half-range Hermite functions \cite{Ball2002}. In Appendix
\ref{sec:radial-method} this is laid out in some detail.

Equation~(\ref{eq:radial-eqn}) is a one-dimensional equation, so
there will be no degeneracy in the eigenvalues $\mu_{m,n}$ for fixed
$m$. In particular, the eigenvalues as function of the interaction
strength $\lambda$ will not cross, and will be continuous functions of
$\lambda$. We thus have $\mu_{m,n}<\mu_{m,n+1}$ for all $n$, where $n$
is the nodal quantum number.

At $\lambda=0$ we regain the harmonic oscillator eigenvalues $2n + |m|
+ 1$. Correspondingly, the eigenfunctions $\psi_{m,n}(r,\theta)$ approaches the
Fock-Darwin orbitals $\varphi_{m,n}(r,\theta)$, i.e., the harmonic
oscillator eigenfunctions. For the radial part,
\[ \lim_{\lambda=0} u_{m,n}(r) = g^{|m|}_n(r) := \sqrt{2} r^{|m|}\tilde{L}_n^{|m|}(r^2)e^{-r^2/2}. \]

Reintroducing spin, the full eigenfunctions
$\Psi=\Psi_{n_1,m_1,n_2,m_2}$ are on the form
\[ \Psi(x_1,x_2) =
\varphi_{n_1,m_1}(\vec{R})\frac{e^{im_2\theta}}{\sqrt{2\pi}}u_{n_2,m_2}(r)
\chi_{s,s_z}, \]
where $s=0$ for odd $m_2$, and $s=1$ for even $m_2$, and $|s_z|\leq s$
is an integer.

Let $R_i = 2n_i + |m_i|$ be the shell numbers for the centre-of-mass
coordinate and relative coordinate, respectively. The eigenvalue
$E=E_{n_1,m_1,n_2,m_2}$ is
\[ E = R_1 + 1 + \mu_{n_2,m_2} , \]
with limit
\[ E \underset{\lambda\rightarrow
  0}{\longrightarrow}  R_1+R_2 + 2, \]
which is the harmonic oscillator eigenvalue.

As the centre-of-mass coordinate transformation conserves harmonic
oscillator energy, at $\lambda=0$, the eigenfunctions that are in the
model space are exactly those obeying $R_1+R_2\leq R$. Turning on the
interaction adiabatically, the eigenpairs we must choose for the
effective Hamiltonian at a given $\lambda$ are exactly
those with $R_1+R_2\leq R$.

The model-space projection $P\Psi$ needed in Eqns.~(\ref{eq:minimize})
and (\ref{eq:minimize2}) is now given by
\[ P\Psi(x_1,x_2) =
\varphi_{n_1,m_1}(\vec{R})\frac{e^{im_2\theta}}{\sqrt{2\pi}}
\left[\tilde{P}^{|m_2|}_{R-R_1} u_{n_2,m_2}(r)\right]\chi_{s,s_z}, \]
where 
\begin{equation} 
  \tilde{P}^{|m|}_R := \sum_{n=0}^{\bar{n}} \ket{g^{|m|}_n}\bra{g^{|m|}_n},
  \quad \bar{n} = \left\lfloor \frac{R - |m|}{2}\right\rfloor, \label{eq:r-projector}
\end{equation}
where $\lfloor x \rfloor$ is the integer part of $x$.  This operator
thus projects onto the $\bar{n}+1$ first radial basis functions with
given $|m|$.

Due to Eqn.~(\ref{eq:Heff-commutator}), the unitary operator $Z$ can be decomposed
into its action on blocks defined by tuples of $n_1,m_1$ and $m_2$
\cite{Kvaal2008a}. The minimization (\ref{eq:minimize}) can then
be applied on block-per-block basis as well. Each sub-problem is
equivalent to the calculation of an effective Hamiltonian
$K_\text{eff}$ of the radial problem for a given $m_2$ and $\bar{n}$.

To this end, let $\bar{n}$ and $m=m_2$ be given. Let $U$ be the $(\bar{n}+1)\times(\bar{n}+1)$ matrix
whose elements are given by
\[ U_{n,k} = \braket{g_n^{|m|}|u_{m,n}}, \quad 0\leq n,k \leq \bar{n}, \] i.e., the model
space projections of the exact eigenvectors with the lowest
eigenvalues. Let $U = X\Sigma Y^\dag$ be the singular value decomposition,
and let $V = XY^\dag$. Then, 
\[ K_\text{eff} = V \diag(E_0,\cdots,E_{\bar{n}}) V^\dag \]
and
\[ 
  \tilde{C}^{\bar{n},|m|} := K_\text{eff} -
  \diag(|m|+1,2+|m|+1,\cdots,2\bar{n}+|m|+1) 
\]
is the $(n_1,m_1,m_2)$-block of the effective interaction.  If we return to
Eqn.~(\ref{eq:interaction-elems}), the effective interaction matrix
elements $\tilde{u}^{ab}_{cd}$ are now given by replacing the matrix
elements $C^{|p-q|}_{n,n+s}$ by the matrix elements
$\tilde{C}^{\bar{n},|p-q|}_{n,n+s}$, where
\[ \bar{n} = \left\lfloor \frac{R - R_1 - |m|}{2} \right\rfloor, \quad
R_1 = N + M - (p+q), \]
where $N$, $M$, $p$ and $q$ are defined immediately after
Eqn.~(\ref{eq:interaction-elems}).

\section{Code organization and use}
\label{sec:code}

\subsection{Overview}
\label{sec:overview}

The main program is called \textsc{qdot}, and processes a
textual configuration file with problem parameters before proceeding
with the diagonalization of the Hamiltonian. Eventually, it writes the
resulting data to a \textsc{Matlab/Gnu Octave} compatible script for
further processing.

As a C++ library as well as stand-alone application, \textsc{OpenFCI}
is organized in several namespaces, which logically separate
independent units. There are three main namespaces: \texttt{manybody},
\texttt{gauss}, and \texttt{quantumdot}. Put simply, \texttt{manybody}
provides generic tools for many-body calculations, such as occupation
number formalism, Slater determinants and CSFs, while \texttt{gauss}
provides tools for orthogonal polynomials and Gaussian
quadrature. These namespaces are independent of each other, and are in
no way dependent on the particular quantum dot model. On the other
hand, \texttt{quantumdot} synthesizes elements from the two former
into a quantum dot FCI library. In \textsc{qdot}, the main work is
thus processing of the configuration file.

Two other namespaces are also defined, being \texttt{simple\_sparse}
and \texttt{simple\_dense}, which are, respectively, simple
implementations of sparse and dense matrices suitable for our
needs. We will not go into details in the present article.

It should be clear that extending and customizing \textsc{qdot} is a
relatively easy task. The application \textsc{qdot} is provided as a
tool with a minimum of functionality, and the interested will almost
certainly desire to further develop this small application.

In order to help with getting started on such tasks, some stand-alone
demonstration applications are provided, all based on the core classes
and functions. These include an interaction matrix element tabulator
\textsc{tabulate}, and a simple program \textsc{pairing} for studying
the well-known pairing Hamiltonian \cite{Richardson1964}, which we
will not discuss further here. Finally, there is a small interactive
console-based Slater determinant demonstration program
\textsc{slater\_demo} as well. These applications will also serve as
indicators of the flexibility of \textsc{OpenFCI}.

\textsc{OpenFCI} does not yet support parallel computation on clusters
of computers, using for example the Message Passing Interface
\cite{MPI}. Future versions will almost certainly be parallelized, but
the present version in fact competes with parallel implementations of
the standard FCI method with respect to convergence due to the
effective interaction implemented, see Sec.~\ref{sec:results}. The
simple structure of \textsc{OpenFCI} also allows users with less
resources to compile and run the code.

\subsection{Core functionality}

The \texttt{manybody} namespace currently contains four main
classes: \texttt{Slater}, \texttt{CsfMachine}, \texttt{NChooseKBitset}, and
\texttt{MatrixMachine}. These will probably form the backbone of any
manybody computation with \textsc{OpenFCI}. 

The class \texttt{Slater} provides Slater determinants, creation and
annihilation operators, and so on. It is based on the standard
template library's (STL) \texttt{bitset} class, which provides generic
bit set manipulations. The class \texttt{NChooseKBitset} provides
means for generating sets of $k$ objects out of $n$ possible
represented as bit-patterns, i.e., bit patterns corresponding to
Slater determinants in the basis $\mathcal{B}$ or $\mathcal{B}'$. This
results in a STL \texttt{vector<Slater>} object, which represent
Slater determinant bases in \textsc{OpenFCI}.

The class \texttt{CsfMachine} is a tool for converting
a basis of Slater determinants into a basis of configurational state
functions. These are represented as \texttt{vector<csf\_block>}
objects, where \texttt{csf\_block} is a \texttt{struct} containing
a few CSFs associated with the same
set of Slater determinants \cite{Rontani2006}.

A CSF basis is again input for the class \texttt{MatrixMachine}, which
is a template class, and generates a sparse matrix $PAP$ of an
operator $A$, where $P$ projects onto the basis. It also handles bases
of pure Slater determinants as they are trivially dealt with in the
CSF framework. The template parameter to \texttt{MatrixMachine} is a
class that should provide the matrix elements $h^{\alpha}_{\beta}$,
$u^{\alpha\beta}_{\gamma\delta}$, etc, of the generic operator given
by
\begin{eqnarray*} 
  A &=& \sum_{\alpha\beta} h^\alpha_\beta a^\dag_\alpha a_\beta +
  \frac{1}{2}\sum_{\alpha\beta\gamma\delta}
  u^{\alpha\beta}_{\gamma\delta} a^\dag_\alpha a^\dag_\beta a_\delta
  a_\gamma \\ & & + \frac{1}{3!}\sum_{\alpha\beta\cdots}
  v^{\alpha\beta\gamma}_{\delta\epsilon\zeta} a^\dag_\alpha a^\dag_\beta
  a^\dag_\gamma a_\zeta a_\epsilon a_\delta .  \notag 
\end{eqnarray*}
Notice, that the indices are generic orbitals, and not assumed to be
on the form $(a,\sigma)$ as in Eqn.~(\ref{eq:H-occ}). 

Currently, only one-, two-, and three-body operators are
implemented. The reason is, that the matrix elements are \emph{not}
computed by directly applying the sum of creation- and annihilation
operators to Slater determinants, since this approach, however
natural, is very inefficient. Instead, we apply Wick's theorem
directly \cite{Raimes1972} on the matrix elements known to be not
identically zero.

In the \texttt{gauss} namespace, several functions are defined which
computes sequences of orthogonal polynomials via recurrence relations
and weights and abscissa for Gaussian quadratures based on these. The
latter is done using the Golub-Welsh algorithm, which only depends on
being able to compute the coefficients of the recurrence relation
\cite{Golub1969}. The most important functions are perhaps
\texttt{computeLaguerrePolys()} and
\texttt{computeGenHalfGaussHermite()}, which computes a sequence of
generalized Laguerre polynomials evaluated at a given set of points
and quadrature rules for generalized half-range Hermite functions,
respectively.

Finally, the \texttt{quantumdot} namespace defines classes and
functions that combined define the quantum dot problem. The class
\texttt{RadialPotential} encapsulates potentials on the form
(\ref{eq:pot1}). It also computes effective
interaction blocks $\tilde{C}^{\bar{n},|m|}$. The class
\texttt{QdotHilbertSpace} provides means for generating the bases
$\mathcal{B}$ and $\mathcal{B}'$, utilizing conservation of angular
momentum, using a fast, custom made algorithm independent of
\texttt{NChooseKBitset}. The class \texttt{QdotFci} sews everything together
and is basically a complete solver for the FCI method with effective
interactions.

\subsection{Sample runs}
\label{sec:results}

\begin{figure}
\includegraphics{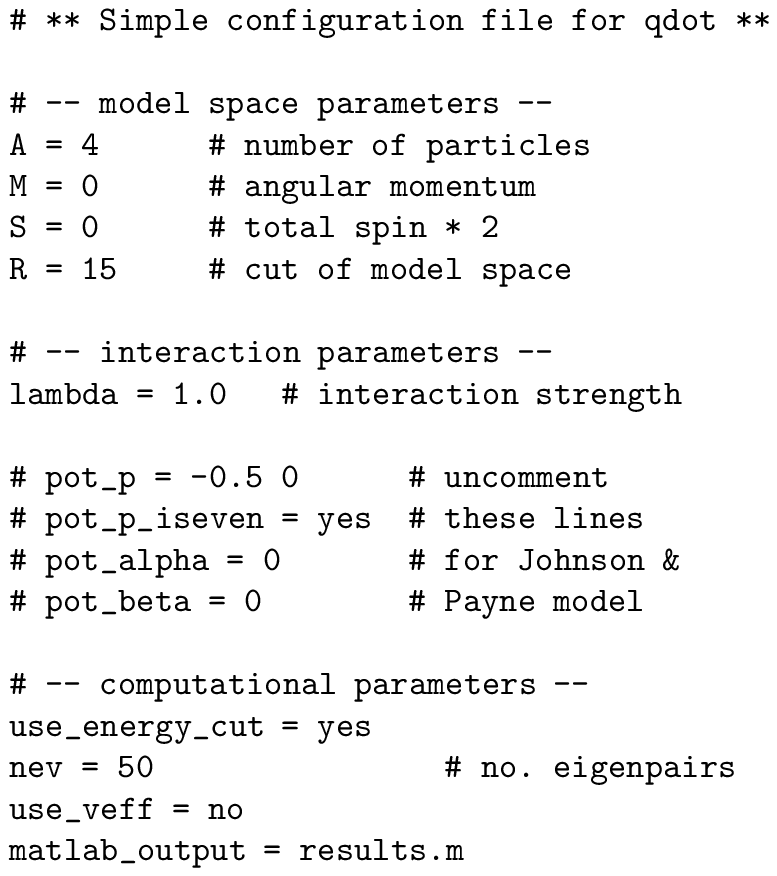}




\caption{Simple configuration file for \textsc{qdot}\label{fig:config}}
\end{figure}

A basic configuration file for \textsc{qdot} is shown in
Fig.~\ref{fig:config}. Varying the parameters \texttt{lambda},
\texttt{R}, \texttt{S} and the number of particles \texttt{A} \footnote{In nuclear physics, it is common to denote
by $A$ the number of particles, i.e., the atomic number. The choice of
the variable name was chosen also partly because $N$ is used
frequently for other purposes in the code.}, and running \textsc{qdot} each time, we
produce a table of ground state energies, shown in Table
\ref{tab:results}. By changing the parameter
\texttt{use\_veff} we turn on and off the effective interaction. The
corresponding effective interaction ground states are also shown in
the table. Notice, that with the effective interaction we obtain the
same precision as the bare interaction, but with much smaller model
spaces. This indicates that \textsc{OpenFCI} can produce results that
in fact compete with parallel implementations of the standard FCI method, even
in its present serial form.

\begin{table}
\caption{Some ground state eigenvalues produced by
  \textsc{qdot} for $N=3,4$ electrons with $\lambda=2$. Both the bare
  and the effective interaction are used\label{tab:results}}
\begin{ruledtabular}
\begin{tabular}{d|d|d|d|d}
\multicolumn{1}{c}{} \vline & \multicolumn{2}{c}{$N=3$, $M=0$,
  $s=\frac{1}{2}$} \vline &
 \multicolumn{2}{c}{$N=4$, $M=s=0$} \\ 
\hline
\multicolumn{1}{c}{$R$} \vline & \multicolumn{1}{c}{$E_{0}$} &
\multicolumn{1}{c}{$E_{0,\text{eff}}$} \vline & \multicolumn{1}{c}{$E_{0}$} &
\multicolumn{1}{c}{$E_{0,\text{eff}}$} \\ \hline
   6 &  9.02370 &  8.96523 & 13.98824 & 13.88832 \\
  10 &  8.97698 &  8.95555 & 13.86113 & 13.83280 \\
  14 &  8.96800 &  8.95465 & 13.84491 & 13.82848 \\
  18 &  8.96411 &  8.95444 & 13.83923 & 13.82761 \\
  22 &  8.96191 &  8.95435 & 13.83626 & 13.82730 \\
  26 &  8.96049 &  8.95430 &          &  \\
  30 &  8.95950 &  8.95428 &          & 
\end{tabular}
\end{ruledtabular}
\end{table}

In Table~\ref{tab:compare} we compare the ground state energies
reported in Ref.~\cite{Rontani2006} with the alternative model space
$\mathcal{P}'$ to the corresponding values produced by \textsc{qdot},
also using $\mathcal{P}'$. This Table also appears in
Ref.~\cite{Kvaal2008}, and serves as a check of the validity of the
calculations.

\begin{table*}
\caption{Comparison of current code and
  Ref.~\cite{Rontani2006}, taken from Ref.~\cite{Kvaal2008}\label{tab:compare}}
\begin{ruledtabular}
\begin{tabular}{c|ccc|dd|dd|dd}
 &  &  &  & \multicolumn{2}{c}{$R=5$} &
  \multicolumn{2}{c}{$R=6$} & \multicolumn{2}{c}{$R=7$} \\
$N$ & $\lambda$ & $M$ & $2s$ &\multicolumn{1}{c}{Current} & \multicolumn{1}{c}{Ref.~\onlinecite{Rontani2006}}
&\multicolumn{1}{c}{Current}  & \multicolumn{1}{c}{Ref.~\onlinecite{Rontani2006}}
&\multicolumn{1}{c}{Current}  & \multicolumn{1}{c}{Ref.~\onlinecite{Rontani2006}} \\
\hline
2 & 1 & 0 & 0 & 3.013626 &        & 3.011020 &        & 3.009236 &      \\
  & 2 & 0 & 0 & 3.733598 & 3.7338 & 3.731057 & 3.7312 & 3.729324 &3.7295\\
  &   & 1 & 2 & 4.143592 & 4.1437 & 4.142946 & 4.1431 & 4.142581 &4.1427\\
3 & 2 & 1 & 1 & 8.175035 & 8.1755 & 8.169913 &        & 8.166708 &8.1671\\
  & 4 & 1 & 1 & 11.04480 & 11.046 & 11.04338 &        & 11.04254 &11.043\\
  &   & 0 & 3 & 11.05428 & 11.055 & 11.05325 &        & 11.05262 &11.053\\
4 & 6 & 0 & 0 & 23.68944 & 23.691 & 23.65559 &        & 23.64832 &23.650\\
  &   & 2 & 4 & 23.86769 & 23.870 & 23.80796 &        & 23.80373 &23.805\\
5  & 2  & 0 & 5 & 21.15093 & 21.15  & 21.13414 & 21.13  & 21.12992        &21.13 \\
  & 4  & 0 & 5 & 29.43528 & 29.44  & 29.30898 & 29.31  & 29.30251 &29.30\\
\end{tabular}   
\end{ruledtabular}
\end{table*}

By uncommenting the lines following the definition of \texttt{lambda},
we override the default Coulomb interaction, and produce a
configuration file for the analytically solvable model given by
Johnson and Payne \cite{Johnson1991}, where the Coulomb interaction is
replaced by the parabolic interaction
\[ U(r_{12}) = -\frac{1}{2}\lambda r_{12}^2. \]
If $\lambda$ is sufficiently small, all the eigenvalues of this model
are on the form
\begin{equation}
  E_{j,k} = 1 + j + (k + N - 1)\sqrt{1 - N\lambda}, \quad j,k\geq 0.
  \label{eq:johnson-eigenvalues}
\end{equation}
Since the potential is smooth, the eigenfunctions are all smooth,
implying exponential convergence with respect to $R$
\cite{Kvaal2008}. We therefore expect very accurate eigenvalues even
with moderate $R$.  In Table \ref{tab:results-johnson} we show the
first eigenvalues along with the error computed for $N=4$
electrons with $\lambda = 1/8$. The computations are done in the
$M=s=s_z=0$ model space with $R=10$ and $R=15$. Some duplicates exist,
and they are included for illustration purposes. It is evident, that
the eigenvalues become very accurate with increasing $R$; a clear
indication of the correctness of the implementation.

\begin{table}
\caption{Results from diagonalizing the Johnson and Payne
  model. Many digits are included due to comparison with exact
  results and high precision\label{tab:results-johnson}} 
\begin{ruledtabular}
\begin{tabular}{c|c|c|c}
\multicolumn{2}{c}{$R=10$} & \multicolumn{2}{c}{$R=15$} \\ \hline
 \multicolumn{1}{c}{$E$} & \multicolumn{1}{c}{$\Delta E$} \vline&
\multicolumn{1}{c}{$E$} & \multicolumn{1}{c}{$\Delta E$} \\
\hline
4.535550207816 & $1.63\cdot 10^{-5}$ & 4.535533958447 & $5.25\cdot 10^{-8}$ \\
5.950417930316 & $6.70\cdot 10^{-4}$ & 5.949751427847 & $3.96\cdot 10^{-6}$ \\
5.950417930316 & $6.70\cdot 10^{-4}$ & 5.949751427847 & $3.96\cdot 10^{-6}$ \\
5.950417930316 & $6.70\cdot 10^{-4}$ & 5.949751427847 & $3.96\cdot 10^{-6}$ \\
5.951592166603 & $1.84\cdot 10^{-3}$ & 5.949760599290 & $1.31\cdot 10^{-5}$ \\
6.243059891817 & $4.19\cdot 10^{-4}$ & 6.242642740293 & $2.05\cdot 10^{-6}$ \\
6.243059891817 & $4.19\cdot 10^{-4}$ & 6.242642740293 & $2.05\cdot 10^{-6}$ \\
6.535776573577 & $2.43\cdot 10^{-4}$ & 6.535534873729 & $9.68\cdot 10^{-7}$ \\
6.535776573577 & $2.43\cdot 10^{-4}$ & 6.535534873729 & $9.68\cdot 10^{-7}$ \\
6.535776573577 & $2.43\cdot 10^{-4}$ & 6.535534873729 & $9.68\cdot 10^{-7}$ \\
7.375904323762 & $1.19\cdot 10^{-2}$ & 7.364103882564 & $1.43\cdot 10^{-4}$ \\
7.375904323762 & $1.19\cdot 10^{-2}$ & 7.364103882564 & $1.43\cdot 10^{-4}$ \\
7.375904323762 & $1.19\cdot 10^{-2}$ & 7.364103882564 & $1.43\cdot 10^{-4}$ \\
7.375904323762 & $1.19\cdot 10^{-2}$ & 7.364103882564 & $1.43\cdot 10^{-4}$ \\
7.375904323762 & $1.19\cdot 10^{-2}$ & 7.364103882564 & $1.43\cdot 10^{-4}$ \\
7.393706556283 & $2.97\cdot 10^{-2}$ & 7.364440927813 & $4.80\cdot 10^{-4}$ \\
7.393706556283 & $2.97\cdot 10^{-2}$ & 7.364440927813 & $4.80\cdot 10^{-4}$ \\
7.393706556283 & $2.97\cdot 10^{-2}$ & 7.364440927813 & $4.80\cdot 10^{-4}$ \\
7.410720999386 & $4.68\cdot 10^{-2}$ & 7.364876152101 & $9.15\cdot 10^{-4}$ \\
7.665921446569 & $9.07\cdot 10^{-3}$ & 7.656945606956 & $9.14\cdot 10^{-5}$
\end{tabular}
\end{ruledtabular}
\end{table}

\section{Conclusion and outlook}
\label{sec:conclusion}

We have presented \textsc{OpenFCI}, an open source full configuration
interaction implementation for quantum dots and similar
systems. \textsc{OpenFCI} also implements a renormalized effective
interaction widely used in nuclear no-core shell model calculations,
and we demonstrated that such interactions are indeed useful in
the quantum dot calculations as well.

\textsc{OpenFCI} is easy to extend and adapt. Possible applications
are computations on systems with more general symmetry-breaking
geometries and in $d=3$ spatial dimensions. Also, a generalization of
the CSF part of the code to handle isobaric spin would allow us to
handle nuclear systems.

There is one more symmetry of the Hamiltonian $H$ that can be
exploited, namely that of conservation of centre-of-mass motion, which
would further reduce the block sizes of the matrices. We exploited
this symmetry for the effective interaction, but it is a fact that it
is a symmetry for the full Hamiltonian as well. Using the energy cut
model space $\mathcal{P}$ we may take care of this symmetry in a way
similar to the CSF treatment \cite{Wensauer2004}.

As mentioned, we have not parallelized the code at the time of
writing, but it is not difficult to do so. A future version will
almost certainly provide parallelized executables, for example using
the Message Passing Interface \cite{MPI}.

\appendix
\section{Centre of mass transformation}
\label{sec:com}

\subsection{Cartesian coordinates}
\label{sec:com-cartesian}

In this appendix, we derive the centre-of-mass (COM) transformation
utilized in Eqn.~(\ref{eq:interaction-elems}) for the interaction
matrix elements $u^{ab}_{cd}$.

The one-dimensional harmonic oscillator (HO) Hamiltonian $(p_x^2+x^2)/2$ is easily diagonalized to yield
eigenfunctions on the form 
\begin{eqnarray}
  \phi_n(x) &=& (2^n n! \sqrt{\pi})^{-1/2} H_n(x) e^{-x^2/2} \notag \\
  &=& (n!)^{-1/2} A_x^n \phi_0(x),
\end{eqnarray}
where $A_x := (x - ip_x)/\sqrt{2}$ is the raising operator in the
$x$-coordinate, and where $\phi_0(x) = \pi^{-1/4}\exp(-x^2/2)$. The
eigenvalues are $n+1/2$.

Using separation of variables, the two-dimensional HO $H_0$ in $x_1$ and
$x_2$ is found to have eigenfunctions on the form
$\Phi_{n_1,n_2}(x_1,x_2):=\phi_{n_1}(x_1)\phi_{n_2}(x_2)$ and eigenvalues $n_1+n_2+1$. Define
the raising operators $A_{x_i} := (x_i - ip_{x_i})/\sqrt{2}$, so
that
\begin{eqnarray}
  \Phi_{n_1,n_2}(\vec{x}) &:=& (n_1!n_2!)^{-1/2}A_{x_1}^{n_1}A_{x_2}^{n_2}
  \Phi_{0,0}(\vec{x}),
  \label{eq:two-dim-hermite}
\end{eqnarray}
where the non-degenerate ground state is given by
\begin{equation} 
  \Phi_{0,0}(x_1,x_2) = \frac{1}{\sqrt{\pi}} e^{ - (x_1^2 + x_2^2)/2 }. 
\end{equation}
Note that this can equally well describe two (distinguishable and
spinless) particles in one dimension.

To this end, we introduce normalized COM frame coordinates by
\begin{equation}
  \begin{bmatrix} \xi_1 \\ \xi_2 \end{bmatrix} =
  \frac{1}{\sqrt{2}}\begin{bmatrix} 1 & 1 \\ 1 & -
    1 \end{bmatrix} \begin{bmatrix} x_1 \\ x_2 \end{bmatrix} =:
  F\begin{bmatrix} x_1 \\ x_2 \end{bmatrix} 
  \label{eq:com-coords}
\end{equation}
The matrix $F$ is symmetric and orthogonal, i.e., $F^TF = F^2 =
1$, transforming a set of Cartesian coordinates into another. The
operator $H_0$ is invariant under this transformation, so the
eigenfunctions have the same form with respect to these coordinates,
viz,
\begin{eqnarray}
  \Phi'_{n_1,n_2}(\xi_1,\xi_2) &:=& \phi_{n_1}(\xi_1)\phi_{n_2}(\xi_2)
  \notag \\
 &=& (n_1!n_2!)^{-1/2}A_{\xi_1}^{n_1}A_{\xi_2}^{n_2}
  \Phi'_{0,0},
\end{eqnarray}
where $A_{\xi_i} = (\xi_i - i p'_i)/\sqrt{2}$ are the raising operators
with respect to the  COM coordinates, and $p'_i$ are the corresponding
momentum components. 

Define the operator $T$ by
\begin{equation}
  T\psi(x_1,x_2) := \psi(\xi_1, \xi_2) =
  \psi\left(\frac{x_1+x_2}{\sqrt{2}},\frac{x_1-x_2}{\sqrt{2}}\right), 
\end{equation}
so that 
\begin{equation}
  T \Phi_{n_1, n_2} := \Phi'_{n_1, n_2}.
\end{equation}
Since $T$ maps eigenfunctions in the two frames onto each other, $T$
must be a unitary operator, and the invariance of $H_0$ under the
coordinate transformation is the same as $[H_0,T]=0$, i.e.,
that energy is conserved. This in turn means that $T$ is block
diagonal with respect to each shell $R=n_1+n_2$, viz,
\begin{eqnarray}
  T \Phi_{R-n_2,n_2}
  &=& \sum_{n = 0}^{R} \braket{\Phi_{R - n, n} | \Phi'_{R-n_2, n_2}} \Phi_{R-n,n}
  \notag \\
  &=:& \sum_{n = 0}^{R} T^{(R)}_{n_2,n} \Phi_{R-n,n},
\end{eqnarray}
where $T^{(R)}$ is the $(R+1)\times(R+1)$ transformation matrix within
shell $R$. It is real, symmetric, and orthogonal. Numerically, the
matrix elements are conveniently computed using two-dimensional
Gauss-Hermite quadrature of sufficiently high order, producing exact
matrix elements.

In a two-dimensional setting, the two-particle harmonic oscillator
becomes a $4$-dimensional oscillator. Let $\vec{r}_i=(x_i,y_i)$,
$i=1,2$, be the particles' coordinates, and let $a=(m_1,n_2)$ and
$b=(m_2,n_2)$ to compress the notation a little. An
eigenfunction is now on the form
\begin{eqnarray}
 \Psi_{a,b}(\vec{r}_1,\vec{r}_2) &:=&
 \Phi_a(\vec{r}_1)\Phi_b(\vec{r}_2) \notag \\
&=& C A_{x_1}^{m_1} A_{y_1}^{n_1}
A_{x_2}^{m_2} A_{y_2}^{n_2} \Psi_{0,0}(\vec{r}_1,\vec{r}_2)
\label{eq:with-ladders}
\end{eqnarray}
where $C = (m_1!n_1!m_2!n_2!)^{-1/2}$.

The COM coordinate transformation now acts in the $x$ and $y$
directions separately, viz, $F$ acts on $x_i$ and $y_i$ to yield the
COM coordinates $\xi_i$ and $\eta_i$: $[\xi_1,\xi_2]^T = F
[x_1,x_2]^T$ and $[\eta_1,\eta_2]^T = F [y_1,y_2]^T$. The induced
operator $T$ again conserves energy.  Let $M = m_1 + m_2$ and $N =
n_1+n_2$. It is readily verifiable that the COM frame transformation
becomes
\begin{eqnarray}
 &&T \Psi_{M-m_2,N-n_2,m_2,n_2} := \Psi'_{M-m_2,N-n_2,m_2,n_2} \notag \\ &=& \sum_{p=0}^{M}
 T^{(M)}_{m_2,p} \sum_{q=0}^{N} T^{(N)}_{n_2,p} \Psi_{M-p,N-q,p,q}. \label{eqn:complete-com}
\end{eqnarray}
Note that the shell number is $R=N+M$, which is conserved by $T$.

\subsection{Centre of mass transformation for Fock-Darwin orbitals}

Consider a Fock-Darwin orbital $\varphi_{n,m}(\vec{r})$ in shell
$R=2n+|m|$ with energy $R+1$. It is straightforward but somewhat
tedious to show that these can be written in terms of co-called
circular raising operators $B_+$ and $B_-$ \cite{Mota2002} defined by
\begin{equation}
  \begin{bmatrix} B_+ \\ B_- \end{bmatrix}
  := \frac{1}{\sqrt{2}}\begin{bmatrix}1 & i \\ 1 & -i \end{bmatrix} \begin{bmatrix} A_x \\ A_y  \end{bmatrix}.
\end{equation}
Letting $\mu = n + \max(0,m)$ and $\nu = n + \max(0,-m)$ (which gives
$\mu,\nu\geq 0$) one obtains
\begin{equation}
  \varphi_{n,m}(\vec{r}) = (\mu!\nu!)^{-1/2} B_+^\mu B_-^\nu
  \Phi_{0,0}(\vec{r}), 
\end{equation}
which should be compared with Eqn.~(\ref{eq:two-dim-hermite}).
Moreover, $R = \mu + \nu$ and $m = \mu - \nu$, giving energy and
angular momentum, respectively. We comment that this is the reason for
the non-standard factor $(-1)^n$ in the normalization of the
Fock-Darwin orbitals in Eqn.~(\ref{eq:fock-darwin-def}).

Let a two-particle HO state be given by
\begin{eqnarray}
  \tilde{\Psi}_{\mu_1,\nu_1,\mu_2,\nu_2} &:=&
  \varphi_{n_1,m_1}(\vec{r}_1)\varphi_{n_2,m_2}(\vec{r}_2), \notag \\
  &=& C B_{1+}^{\mu_1} B_{1-}^{\nu_1} B_{2+}^{\mu_2} B_{2-}^{\nu_2} \Psi_{0,0}(\vec{r}_1,\vec{r}_2),
  \label{eq:fock-darwin-2}
\end{eqnarray} 
where $\mu_i=n_i+\max(0,m_i)$ and $\nu_i=n_i + \max(0,-m_i)$, and
where $C=(\mu_1!\nu_1!\mu_2!\nu_2!)^{-1/2}$.
We will now prove that, in fact, when applying the centre-of-mass
transformation to Eqn.~(\ref{eq:fock-darwin-2}), we obtain an
expression on the same form as Eqn.~(\ref{eqn:complete-com})
viz,
\begin{eqnarray}
 &&T \tilde{\Psi}_{M-\mu_2,N-\nu_2,\mu_2,\nu_2} := \tilde{\Psi}'_{M-\mu_2,N-\nu_2,\mu_2,\nu_2} \notag \\ &=& \sum_{p=0}^{M}
 T^{(M)}_{\mu_2,p} \sum_{q=0}^{N} T^{(N)}_{\nu_2,q}
 \tilde{\Psi}_{M-p,N-q,p,q}, \label{eqn:complete-com-circ}
\end{eqnarray}
where $M := \mu_1+\mu_2$ and $N:=\nu_1+\nu_2$.

To this end, we return to the raising operators $A_{\xi_i}$ and
$A_{\eta_i}$, and express them in terms of $A_{x_i}$ and $A_{y_i}$. By
using Eqn.~(\ref{eq:com-coords}), we obtain
\begin{equation}
  \begin{bmatrix} A_{\xi_1} \\ A_{\xi_2} \end{bmatrix} =
  F \begin{bmatrix} A_{x_1} \\ A_{x_2} \end{bmatrix}, 
  \label{eq:com-ladder-trafo}
\end{equation}
and similarly for $A_{\eta_i}$ in terms of $A_{y_i}$. In terms of the
raising operators, the COM transformation becomes
\begin{eqnarray}
  \Psi_{m_1,n_1,m_2,n_2}' &=& C \left(A_{x_1} + A_{x_2}\right)^{m_1}
  \left(A_{y_1} + A_{y_2}\right)^{n_1} \notag \\ &\times& \left(A_{x_1} -
    A_{x_2}\right)^{m_2} \left(A_{y_1} - A_{y_2}\right)^{n_2}
  \Psi_{0,0}, \label{eq:com-raising}
\end{eqnarray}
where 
\begin{equation}
  C=(2^{n_1+n_2+m_1+m_2}n_1!n_2!m_1!m_2!)^{-1/2},
\end{equation}
and we have used Eqn.~(\ref{eq:with-ladders}), but in the analogous
COM case. Expanding the powers using the binomial formula (and the
fact that the raising operators commute), we obtain a linear
combination of the individual eigenfunctions, which must be identical
to Eqn.~(\ref{eqn:complete-com}).

Let the COM circular ladder operators be defined by
\begin{equation}
  \begin{bmatrix} B'_{j+} \\ B'_{j-} \end{bmatrix}
  := \frac{1}{\sqrt{2}}\begin{bmatrix}1 & i \\ 1 & -i \end{bmatrix} \begin{bmatrix} A_{\xi_j} \\ A_{\eta_j}  \end{bmatrix}.
\end{equation}
Using Eqn.~(\ref{eq:com-ladder-trafo}), we obtain that the
circular raising operators transform \emph{in the same way} as the
Cartesian operators when going to the COM frame, i.e.,
\begin{eqnarray}
  \begin{bmatrix} B'_{1+} \\ B'_{2+} \end{bmatrix} =
  F \begin{bmatrix} B_{1+} \\ B_{2+} \end{bmatrix},
\end{eqnarray}
and similarly for $B'_{i-}$ and terms of $B_{i-}$. Using
Eqn.~(\ref{eq:fock-darwin-2}) in the COM case, we obtain
\begin{eqnarray}
  \tilde{\Psi}_{\mu_1,\nu_1,\mu_2,\nu_2}' &=& C \left(B_{1+} + B_{2+}\right)^{\mu_1}
  \left(B_{1-} + B_{2-}\right)^{\nu_1} \notag \\ &\times& \left(B_{1+} -
    B_{2+}\right)^{\mu_2} \left(B_{1-} - B_{2-}\right)^{\nu_2}
  {\Psi}_{0,0}, \label{eq:com-raising2}
\end{eqnarray}
with
\begin{equation}
  C=(2^{\mu_1+\nu_1+\mu_2+\nu_2}\mu_1!\nu_1!\mu_2!\nu_2!)^{-1/2}.
\end{equation}
Eqn.~(\ref{eq:com-raising2}) is on the same form as Eqn.~(\ref{eq:com-raising}). Again, by
expanding the powers using the binomial formula (and that the raising
operators commute), we obtain a linear combination with coefficients
identical to those of the expansion of Eqn.~(\ref{eq:com-raising}). It then
follows that Eqn.~(\ref{eqn:complete-com-circ}) holds.

\section{Numerical treatment of radial problem}
\label{sec:radial-method}

We now briefly discuss the numerical method used for solving the
radial problem (\ref{eq:radial-eqn}), i.e., the eigenvalue problem for
the operator $K_{|m|}$ defined in Eqn.~(\ref{eq:radial-ham}). This is
an eigenproblem in the Hilbert space $L^2([0,\infty), r\rmd r)$, where
the measure $r \rmd r$ is induced by the polar coordinate
transformation. The inner product on this space is thus given by
\begin{equation}
 \braket{f|g} = \int_0^\infty f(r)g(r) r\rmd r.
\end{equation}
Let the Fock-Darwin orbitals be given by
\begin{equation}
  \varphi_{n,m}(r,\theta) = \frac{e^{im\theta}}{\sqrt{2\pi}}
g_n^{|m|}(r),
\end{equation}
with radial part
\begin{equation}
  g_n^{|m|}(r) := \sqrt{2}\tilde{L}_n^{|m|}(r^2) r^{|m|} \exp(-r^2/2)
\end{equation}
Thus,
\[ \braket{g_n^{|m|}|g_{n'}^{|m|}} = \delta_{n,n'}, \]
and these functions form an orthonormal sequence in $L^2([0,\infty),
r\rmd r)$ for fixed $|m|$.

In the electronic case, $U(\sqrt{2}r) = \lambda/{\sqrt{2}r}$ has a
singularity at $r=0$ which gives rise to a cusp in the eigenfunction $u_{m,n}(r)$
at $r=0$, or in one of its derivatives. Away from $r=0$, the
eigenfunction is smooth. These considerations are also true for more
general potentials smooth for $r>0$.

Diagonalizing the matrix of $K$ with respect to the truncated basis
$\{g_n^{|m|}(r)\}_{n=0}^{\bar{n}}$ will give eigenpairs converging slowly with respect
to increasing $\bar{n}$ due to the non-smoothness of $u_{n,m}(r)$ at $r=0$. This
is easily seen for the $m=0$ case and $\lambda = 1$, which has the
exact ground state
\[ u_{0,0}(r) = \left(r + \frac{1}{\sqrt{2}}\right) e^{-r^2/2}, \]
a polynomial of odd degree multiplied by a Gaussian. The cusp at $r=0$
is evident. However, 
\[ g_n^0(r) = \sqrt{2}\tilde{L}_n(r^2)e^{-r^2/2}, \]
which are all \emph{even} polynomials. It is clear, that $\bar{n}$
must be large to resolve the cusp of $u_{0,0}(r)$. 

The eigenproblem is best solved using a
basis of generalized half-range Hermite functions $f_j(r)$
\cite{Ball2002}, which will resolve the cusp nicely. These functions are defined by
\[ f_j(r) := P_j(r)\exp(-r^2/2), \]
where $P_j(r)$ are the orthonormal polynomials defined by
Gram-Schmidt orthogonalization of the monomials $r^k$ with respect
to the weight function $r\exp(-r^2)$. Thus,
\[ \braket{f_j|f_{j'}} = \int_{0}^\infty f_j(r)f_{j'}(r) r \rmd
r = \delta_{j,j'}. \] The fundamental difference between $f_j(r)$ and
$g_n^{|m|}(r)$ is that the latter contains \emph{only} even (odd) powers of
$r$ for even (odd) $|m|$. Both sets constitute orthonormal bases, but $f_j(r)$ will in
general have better approximation properties.

Moreover, since $\deg(r^{|m|}L_n^{|m|}(r^2)) = 2n+|m|$,
\begin{equation} 
  g_n^{|m|}(r) = \sum_{j=0}^{2n + |m|} \braket{f_j|g_n^{|m|}}
f_j(r)
\label{eq:an-expansion}
\end{equation}
gives the Fock-Darwin orbitals as a \emph{finite} linear combination of the
generalized half-range Hermite functions, while the converse is not
possible.

Computing the matrix of $K$ with respect to
$\{f_j(r)\}_{j=0}^{\bar{j}}$ and diagonalizing will give eigenpairs
converging exponentially fast with respect to increasing
$\bar{j}$. The resulting eigenfunctions' expansion in $g_n^{|m|}$ are
readily computed using Eqn.~(\ref{eq:an-expansion}), whose
coefficients $\braket{f_j|g^{|m|}_n}$ can be computed numerically
exactly using Gaussian quadrature induced by $P_J(r)$, for $J$
sufficiently large.

The basis size $\bar{j}$ to use in the diagonalization depends on how many eigenfunctions
$\bar{n}$ we desire. We adjust $\bar{j}$ semi-empirically, noting that
$2\bar{n}+|m|$ is sufficient to resolve $g_{\bar{n}}^{|m|}$, and
assuming that the exact eigenfunctions are dominated by the latter. We
then add a fixed number $j_0$ to get $\bar{j} = 2\bar{n} + |m| + j_0$,
and numerical experiments confirm that this produces eigenvalues that
indeed have converged within desired precision.

\section*{Acknowledgments}
\label{sec:ack}

The author wishes to thank Prof.~M.~Hjorth-Jensen (CMA) for helpful
discussions, suggestions and feedback. This work was financed by CMA through
the Norwegian Research Council.


\end{document}